\documentclass[final]{svjour2}
\usepackage{graphicx}
\usepackage{rotating}
\usepackage{amsmath,amssymb}
\usepackage{mathptmx}
\usepackage[numbers]{natbib}
\usepackage{color}

\makeatletter
\journalname{Journal of Low Temperature Physics}

\bibpunct{}{}{,}{s}{}{,}

\begin{document}

\newcommand{\hdblarrow}{H\makebox[0.9ex][l]{$\downdownarrows$}-}
\title{Vortex polygons and their stabilities in Bose-Einstein condensates and field theory}

\author{Michikazu Kobayashi$^1$ \and Muneto Nitta$^2$}

\institute{1:Department of Physics, Kyoto University, Oiwake-cho, Kitashirakawa, Sakyo-ku, Kyoto, 606-8502, Japan\\
\email{michikaz@scphys.kyoto-u.ac.jp}
\\2: Department of Physics, and Research and Education Center for Natural Sciences, Keio University, Hiyoshi 4-1-1, Yokohama, Kanagawa 223-8521, Japan}

\date{\today}

\maketitle


\begin{abstract}

We study vortex polygons and their stabilities in miscible two-component Bose-Einstein condensates,
and find that vortex polygons are stable for the total circulation $Q \leq 5$, metastable for $Q = 6$, and unstable for $Q \geq 7$.
As a related model in high-energy physics, we also study the vortex polygon of the baby-Skyrme model with an anti-ferromagnetic potential term, and compare both results.

\end{abstract}

\section{Introduction}

Quantized vortices have been one of the major topics in superfluid systems and play essential roles to determine the statistical and dynamical properties of the systems.
One of the exotic aspects is stable structure formed by many vortices.
For simple superfluid systems such as superfluid $^4$He and conventional superconductors, the most popular stable state is triangular Abrikosov lattice under the rotation or the magnetic field, respectively.
With a small number of vortices, vortex polygon states and their stabilities have also been studied \cite{Campbell:1979}.
Within the point vortex approximation for quantized vortices, it was reported that vortex monomer ($Q = 1$), dimer ($Q = 2$), regular triangular ($Q = 3$), and square ($Q = 4$) states were stable, where $Q$ is the total topological charge of the system being equivalent to the total number of vortices.
Although the regular vortex pentagon ($Q = 5$) is also stable, there appears the metastable double-ringing vortex state.
For $Q = 6$ and $7$, however, regular vortex hexagon and heptagon become metastable, and regular vortex polygons become unstable for $Q \geq 8$.
Such vortex polygons with $1 \leq Q \leq 7$ have been experimentally observed in superfluid $^4$He \cite{Yarmchuk:1982}.
Vortex polygons have also been studied in hydrodynamics for a long time \cite{fluid}.
In classical fluid with the continuous circulation, vortex polygons with less than seven vortices are shown to be stable or metastable, while those with more than seven are unstable.
One example realized in nature is a vortex hexagon found by Cassini in the north poles of Saturn \cite{saturn}.

Another exotic aspect is vortex molecules studied in multi-component Bose-Einstein condensates (BECs) \cite{2-comp-BEC,Cipriani:2013}, multi-gap superconductors \cite{multi-gap-SC}, superfluid $^3$He \cite{Volovik:2003}, and nonlinear optics \cite{optics}.
In the cases of BECs and superconductors, fractional vortices in different components with fractional circulations can form a various structure of stable and metastable state.
In this paper, we study the vortex polygons formed by fractional vortices and their stabilities in a two-component BEC, where two fractional vortices of each component form a unit topological charge $Q = 1$.
We find that vortex polygons are stable for $Q \leq 4$, metastable for $Q =5, 6$, and unstable for $Q \geq 7$.

Vortex polygons can also be considered in elementary particle physics as models of stable bound states \cite{molecule-FS,Kobayashi:2013}.
As a field theoretical model, we also consider the baby-Skyrme model which is an $O(3)$ nonlinear sigma model with the baby-Skyrme term.
With an anti-ferromagnetic potential term, the model admits vortex molecules and polygons.
In contrast to the two-component BEC, the vortex polygons are stable for $Q \leq 6$ and metastable for $Q \geq 7$.

\section{Vortex polygon in a two-component BEC}

We start with the mean-field Lagrangian density for the two-component BEC in 2-spatial dimension:
\begin{align}
\begin{split}
\mathcal{L}_{\mathrm{BEC}} = \sum_i \bigg\{& \frac{i}{2} \big( \psi_i^\ast \dot{\psi_i} - \dot{\psi_i}^\ast \psi_i \big) - \frac{1}{2} |\nabla \psi_i|^2
 - \frac{g_{ii} \rho}{2} |\psi_i|^4 \\
 & - \frac{g_{12} \rho}{2} |\psi_i|^2 |\psi_{3-i}|^2 - \frac{i \Omega}{2} {\bf x} \times \big( \psi_i^\ast \nabla \psi_i - \psi_i \nabla \psi_i^\ast \big) \bigg\}, \label{eq-Lagrangian-BEC}
\end{split}
\end{align}
where $\psi_i$ ($i = 1, 2$) is the order parameter for the condensate of the $i$-th component, and $\rho$ is the mean particle density $\rho = (1 / L^2) \int d^2x\: \sum_i |\psi_i|^2$, with the system size $L$.
The third and fourth terms in Eq. \eqref{eq-Lagrangian-BEC} are intra- and inter-component interaction energies with the coupling constant $g_{ii}$ and  $g_{12}$ respectively.
The fifth term is the rotation energy with the angular velocity $\Omega$.
Choosing $g_{11} = g_{22} = g$ for simplicity, the third and fourth terms can be rewritten as
\begin{align}
- \frac{(g + g_{12}) \rho}{4} (|\psi_1|^2 + |\psi_2|^2)^2 - \frac{(g - g_{12}) \rho}{4} (|\psi_1|^2 - |\psi_2|^2)^2.
\end{align}
Considering the situation of $g + g_{12} \gg 1$, we can assign the constraint $|\psi_1|^2 + |\psi_2|^2 = 1$ for further simplicity.
The energy density of static configurations is
\begin{align}
\begin{split}
\mathcal{E}_{\mathrm{BEC}} &= \sum_i \bigg\{ \frac{1}{2} |\nabla \psi_i|^2 + \frac{i \Omega}{2} {\bf x} \times \big( \psi_i^\ast \nabla \psi_i - \psi_i \nabla \psi_i^\ast \big) \bigg\} \\
&\phantom{=\ } + \frac{(g - g_{12}) \rho}{4} (|\psi_1|^2 - |\psi_2|^2)^2. \label{eq-energy-BEC}
\end{split}
\end{align}
Under $g > g_{12}$, the miscible state $|\psi_1|^2 = |\psi_2|^2$ is energetically favored as the static configuration.

The ansatz of the vortex polygon state can be given by
\begin{align}
\psi_{1,2} = \frac{1}{\sqrt{2}} \bigg\{ e^{i Q \theta} \cos\frac{f(r)}{2} \pm i \sin\frac{f(r)}{2} \bigg\} \label{eq-polygon-BEC}
\end{align}
in the polar coordinates $(r,\theta)$.
The monotonically decreasing function $f(r)$ satisfies $f(r \to 0) \to \pi$ and $f(r \to \mathrm{boundary}) \to 0$.
At $f(r) = \pi / 2$ and $Q \theta = - \pi / 2$ ($Q \theta = \pi / 2$), there is a half-quantized vortex of the first (second) component with $\psi_1 = 0$ ($\psi_2 = 0$) and $\psi_2 = - i$ ($\psi_1 = i$).
The circulation is given by
\begin{align}
\frac{1}{2 \pi} \sum_i \oint d{\bf l} \times \mathrm{Im} \big[\psi_i^\ast \nabla \psi_i \big] = \frac{1}{2}
\end{align}
around the vortex core.
Totally, there are $Q$ half-quantized vortices for each component, and $Q$ is defined as the topological charge of the system.

\begin{figure}[tbh]
\centering
\includegraphics[width=0.95\linewidth]{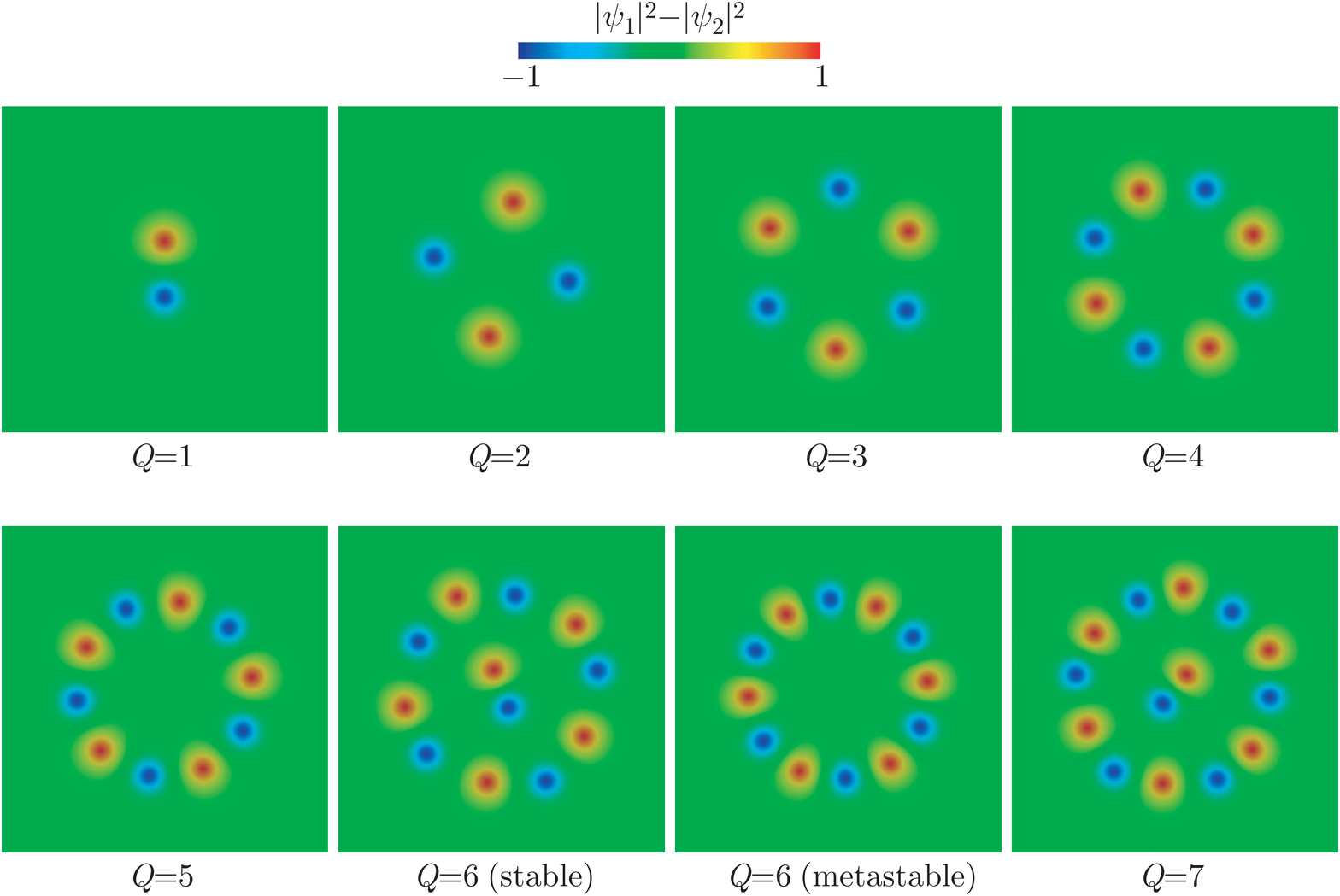}
\caption{\label{fig-polygon-BEC} Stable and metastable vortex states for the energy density \eqref{eq-energy-BEC} with the topological charge $1 \leq Q \leq 7$ in the region $-1 < x, y < 1$. As the numerical parameters, we fix $(g - g_{12}) \rho = 160$, and $\Omega = 0.72 Q$.
}
\end{figure}
By using a relaxation method, we numerically obtain stable and metastable static states for the energy density \eqref{eq-energy-BEC}.
Figure \ref{fig-polygon-BEC} shows $|\psi_1|^2 - |\psi_2|^2$ of all the static states with $1 \leq Q \leq 7$.
We can find the position of vortices of $\psi_1$ ($\psi_2$) component as the minimum (maximum) value of $|\psi_1|^2 - |\psi_2|^2$.
For the unit topological charge $Q = 1$, one can find a pair of half-quantized vortices as a vortex molecule.
For $Q = 2$, two vortex molecules face to each other with opposite orientations.
They constitute a vortex tetragon.
Since the same vortices are placed at diagonal corners, the configuration is $\mathbb{Z}_2$ axisymmetric.
For $Q = 3$, three vortex molecules with six half-quantized vortices constitute a regular hexagon with a $\mathbb{Z}_3$ axisymmetry.
The structures of $Q = 2$ and $3$ resemble those in a vortex lattice found in a two-component BEC under rotation \cite{Cipriani:2013}.
For $Q = 4$ and $5$, the situations is almost the same, {\it i.e.}, one finds that four and five vortex molecules with eight and ten fractional vortices constitute regular octagon and decagon with $\mathbb{Z}_4$ and $\mathbb{Z}_5$ axisymmetries, respectively.
In general, for the topological charge $Q$, we expect $2 Q$ half-quantized vortices to be placed on a circle in a $\mathbb{Z}_Q$ axisymmetric way.
The rotational $SO(2)$ symmetry in the $x$--$y$ plane is spontaneously broken in all cases to a discrete subgroup $\mathbb{Z}_Q$.
For $Q = 6$, however, the vortex dodecagon constituted by twelve half-quantized vortices with a $\mathbb{Z}_6$ axisymmetry becomes metastable.
The stable state has the double-ringing structure of half-quantized vortices: ten half-quantized vortices surround a vortex molecule at the center, forming a decagon (not a regular decagon).
Because of the vortex molecule at the center, the system has no axisymmetry.
For $Q = 7$, the vortex tetradecagon is neither stable nor metastable but unstable, and only double-ringing structures appear as the stable static configurations without any other metastable states.

We compare our result to the case of a scalar BEC \cite{Campbell:1979} in which an integer vortex carries the unit circulation.
For $Q = 1, 2, 3$, and $4$, our results are similar to those for the scalar BEC: $Q$ integer vortices form monomer, dimer, regular triangular, and square with $SO(2)$, $\mathbb{Z}_2$, $\mathbb{Z}_3$, and $\mathbb{Z}_4$ axisymmetries.
For $Q = 6$, the result is also similar: there are the metastable regular vortex hexagon and the stable double-ringing vortices with a vortex at the center and surrounding $5$ vortices.
In contrast to our result, the latter state has $\mathbb{Z}_5$ axisymmetry because there is only one vortex at the center having $SO(2)$ axisymmetry.
For $Q = 5$ and 7, our results rather differ from those for the scalar BEC which has a metastable double-ringing vortex state with a $\mathbb{Z}_4$ axisymmetry for $Q = 5$ and a metastable regular vortex heptagon for $Q = 7$.

\section{Vortex polygon in a baby-Skyrme model}

We next consider an nonlinear $O(3)$ sigma model in 2-spatial dimension described by a three vector of scalar fields ${\bf n} = \begin{pmatrix} n_x & n_y & n_z \end{pmatrix}$ with a constraint ${\bf n} \cdot {\bf n} = 1$.
The Lagrangian density is given by
\begin{align}
\mathcal{L}_{\mathrm{FS}} = \frac{1}{2} {\bf \partial_\mu n} \cdot {\bf \partial^\mu n} - \kappa \big\{ {\bf n} \cdot (\partial_\mu {\bf n} \times \partial_\nu {\bf n}) \big\}^2 - m^2 n_z^2,
\end{align}
with $\mu, \nu = t, x, y$.
The state with $n_z = 0$ and $n_x^2 + n_y^2 = 1$ is energetically favored with the third potential term in the Lagrangian which is known in anti-ferromagnets and the XY model, while the potential term takes the form of $m^2 (1 - n_z^2)$ in ferromagnets and the Ising model.
In the presence of this potential term, a static configuration with vortices is unstable to shrinking from the Derrick's scaling argument \cite{Derrick:1964}, and it can be stabilized in the presence of the second term of the Lagrangian which is known as the baby Skyrme term.
The energy density of static configurations is
\begin{align}
\mathcal{E}_{\mathrm{FS}} = \frac{1}{2} |\nabla {\bf n}|^2 + \kappa \big\{ {\bf n} \cdot (\partial_\mu {\bf n} \times \partial_\nu {\bf n}) \big\}^2 + m^2 n_z^2. \label{eq-energy-FS}
\end{align}
The three vector ${\bf n}$ and the order parameter $\psi_i$ of the 2-component BEC can be related to each other through the Hopf map ${\bf n} = \langle \psi | {\bf \sigma} | \psi \rangle$ with Pauli matrices ${\bf \sigma}$.
With the Hopf map, the ansatz \eqref{eq-polygon-BEC} of the vortex polygon ansatz for the 2-component BEC can be transformed to \begin{align}
n_x = \cos f(r), \quad
n_y = - \sin f(r) \cos(Q \theta), \quad
n_z = \sin f(r) \sin(Q \theta). \label{eq-polygon-FS}
\end{align}
At the position of vortex; $f(r) = \pi/2$ and $Q \theta = - \pi /2$ ($Q \theta = \pi / 2$), the core is filled with $n_z = -1$ ($n_z = 1$), and the state winds counterclockwise (clockwise) within the manifold of the ground state $n_x^2 + n_y^2 = 1$ around the vortex core.
This state coincides with the baby skyrmion ansatz of 2-dimensional $O(3)$-nonlinear sigma model with the topological charge $Q \in \pi_2(S^2) \simeq \mathbb{Z}$ defined as $Q = (1 / 4 \pi) \int d^2x\: {\bf n} \cdot \{ (\partial_x {\bf n}) \times (\partial_y {\bf n}) \}$.

\begin{figure}[tbh]
\centering
\includegraphics[width=0.95\linewidth]{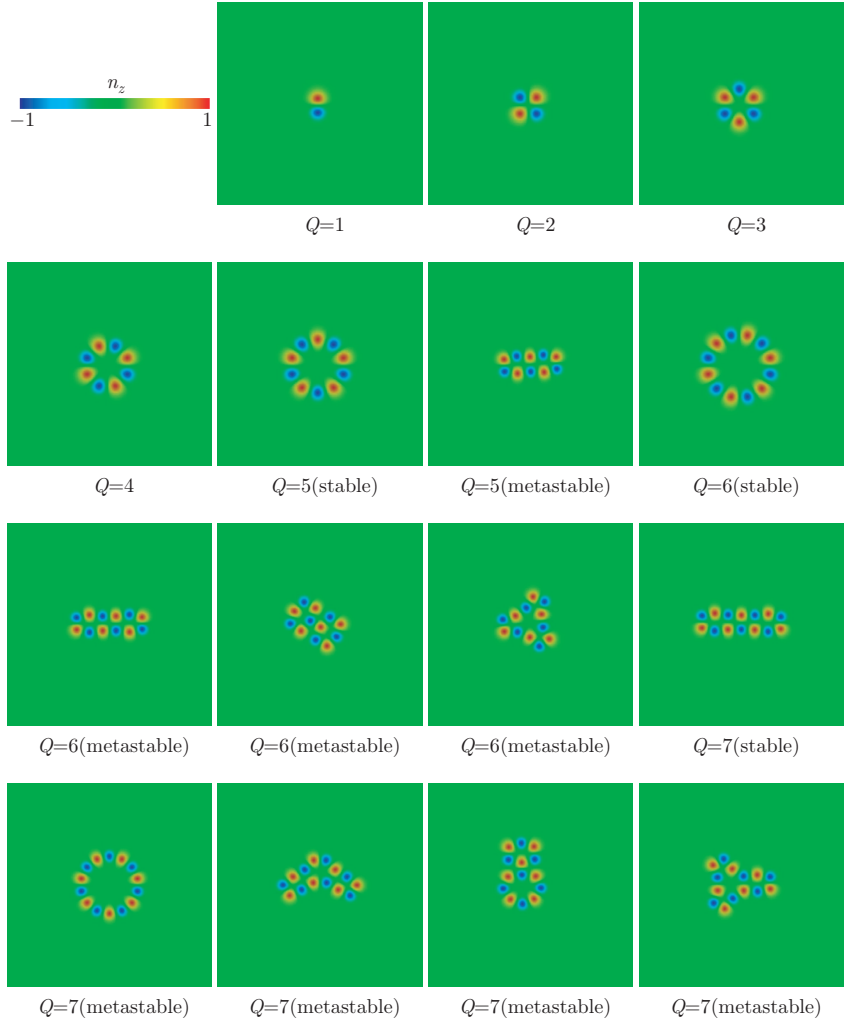}
\caption{\label{fig-polygon-FS} Stable and metastable vortex states for the energy density \eqref{eq-energy-FS} with the topological charge $1 \leq Q \leq 7$ in the region $-1.16 < x, y < 1.16$. As the numerical parameters, we fix $\kappa = 0.002$ and $m^2 = 800$.
}
\end{figure}
We numerically obtain stable and metastable static states for the energy density \eqref{eq-energy-FS} by the relaxation method.
Figure \ref{fig-polygon-FS} shows $n_z$ for all the static states with $1 \leq Q \leq 7$.
As described above, we can find the position of vortices as the minimum and maximum value of $n_z$ which coincides with $|\psi_1|^2 - |\psi_2|^2$ for the 2-component BEC through the Hopf map.
For $Q = 1$, $2$, $3$, and $4$, the results are almost same as those for the 2-component BEC; ground states have $2 Q$ vortices making regular $2 Q$-polygons in a $\mathbb{Z}_Q$ axisymmetric way.
For $Q = 5$, the regular vortex decagon is stable as well as the 2-component BEC, while we have a new metastable vortex arrayed state as shown in Fig. \ref{fig-polygon-FS}.
For $Q = 6$, the regular vortex dodecagon remains stable in spite that the regular vortex dodecagon is metastable for the 2-component BEC.
Besides the vortex arrayed state, we have the other two metastable states which has not been discussed in the previous work \cite{Kobayashi:2013}.
For $Q = 7$, the vortex arrayed state becomes stable and four different metastable states appear.
The regular vortex tetradecagon is included in the metastable states and has the minimum energy among them.

\section{Conclusions}

We have investigated vortex polygons of the half-quantized vortices and their stability in the miscible 2-component BEC and the XY (or anti-ferromagnetic) baby-Skyrme model which has the anti-ferromagnetic (XY) potential $m^2 n_z^2$.
For $1 \leq Q \leq 4$, we obtain only stable regular vortex polygons with a $\mathbb{Z}_Q$ axisymmetry for both models.
For $Q = 5$, besides the stable regular vortex decagon, we have another metastable arrayed vortex structure in the baby-Skyrme model.
For $Q = 6$, the regular vortex dodecagon becomes metastable with the stable double-ringing structure for the 2-component BEC, while the regular vortex dodecagon remains stable for the baby-Skyrme models.
For $Q = 7$, the regular vortex tetradecagon is unstable for the 2-component BEC and metastable accompanied by the stable arrayed vortices and the other 3 metastable states for the baby Skyrme model.
As a result, including the case of vortices in the scalar BEC, the stability of the regular vortex polygon drastically changes in the $5 \leq Q \leq 7$ region, and strongly depends on the models and interactions between vortices.
As our future work, we have to consider the experimental situation of vortex polygons.
In the experimental setup, the BEC is trapped and its trapping geometry also affects the stability and the stable vortex structure.
We will soon report this topic elsewhere.

\begin{acknowledgements}
This work is supported in part by Grant-in-Aid for Scientific Research (Grants No. 22740219 (M.K.) and No. 23740198 and No. 25400268 (M.N.)) and the rowk of M. N. is also supported in part by the ``Topological Quantum Phenomena" Grant-in-Aid for Scientific Research on Innovative Areas (No. 23103515 and No. 25103720) from the Ministry of Education, Culture, Sports, Science and Technology (MEXT) of Japan.
\end{acknowledgements}

\pagebreak

\end{document}